\documentclass[12pt]{iopart}

\usepackage{iopams}
\usepackage{setstack}
\usepackage{amssymb}
\usepackage{graphicx}
\usepackage{cite}

\newcommand{\bra}[1]{\langle #1 |}
\newcommand{\ket}[1]{|#1 \rangle}
\newcommand{\braket}[2]{\langle#1|#2\rangle}

\begin{document}

\title{Variable range of the RKKY interaction in edged graphene}
\author{J. M. Duffy$^{(a)}$, P. D. Gorman$^{(a)}$, S. R. Power$^{(c)}$ and M. S. Ferreira$^{(a, b)}$}
\address{a) School of Physics, Trinity College Dublin, Dublin 2, Ireland \\
b) CRANN, Trinity College Dublin, Dublin 2, Ireland \\
c) Center for Nanostructured Graphene (CNG), DTU Nanotech, Department of Micro- and Nanotechnology,
Technical University of Denmark, DK-2800 Kongens Lyngby, Denmark }

\date{\today}
\pacs{}

\begin{abstract}
The indirect exchange interaction is one of the key factors in determining the overall alignment of magnetic impurities embedded in metallic host materials.
In this work we examine the range of this interaction in magnetically-doped graphene systems in the presence of armchair edges using a combination of analytical and numerical Green function (GF) approaches.
We consider both a semi-infinite sheet of graphene with a single armchair edge, and also quasi-one-dimensional armchair edged graphene nanoribbons (GNRs).
While we find signals of the bulk decay rate in semi-infinite graphene and signals of the expected one-dimensional decay rate in GNRs, we also find an unusually rapid decay for certain instances in both, which manifests itself whenever the impurities are located at sites which are a multiple of three atoms from the edge.
This decay behavior emerges from both the analytic and numerical calculations, and the result for semi-infinite graphene can be interpreted as an intermediate case between ribbon and bulk systems.
\end{abstract}

\maketitle

\section{Introduction}
  Graphene, the two-dimensional carbon allotrope, has been in the scientific limelight for almost a decade now due to a range of fascinating and experimentally realizable physical properties
    \cite{geim_rise_2007, geim_graphene:_2009, pesin_spintronics_2012, kane_quantum_2005,balandin_superior_2008}.
  Interest in this material is further fueled by its potential application in a variety of fields, including spintronics, 
    which attempts to exploit the charge and spin degrees of freedom of electrons in order to develop the next generation of nanoscale devices.
  Graphene is predicted to display very weak spin-orbit and hyperfine interactions which are
    common sources of spin-scattering and decoherence, and thus appears as a promising candidate material for the transport of spin information in such devices\cite{huertas-hernando_spin-orbit_2006,yazyev_hyperfine_2008}.
  
  Fundamental to the field of spintronics is the indirect exchange coupling (IEC) which determines the alignment of magnetic objects in metallic systems. 
  This interaction, mediated by the conduction electrons of the metallic host, makes separate magnetic objects aware of their mutual presence and forces the magnetizations to adopt the most energetically favorable alignment.
  When calculated in the framework of second order perturbation theory, the IEC is commonly known as the Ruderman-Kittel-Kasuya-Yosida (RKKY) interaction 
    \cite{ruderman_indirect_1954,yafet_ruderman-kittel-kasuya-yosida_1987,edwards_oscillations_1991,dalbuquerque_e_castro_theory_1994,ferreira_nature_1994}.
  Despite their slight distinction, both terminologies are often used interchangeably and will be adopted here as equivalent.

  The investigation of the IEC in multilayer systems played a pivotal role in the development of the early spintronic devices, such as giant magnetoresistance stacks and spin valves\cite{baibich_giant_1988,binasch_enhanced_1989,bruno_oscillatory_1991}.
  Recently, experimental progress has been made in measuring the interaction between individual magnetic atoms on surfaces\cite{kaiser_magnetic_2007,meier_revealing_2008,zhou_strength_2010}.
  The basic features of the interaction are well documented, and it is generally seen to oscillate and decay as a function of the separation between the magnetic objects. 
  The oscillation period is closely linked to the Fermi surface of the host material, and the decay rate to the dimensionality of the system.
  In a two dimensional host, the RKKY interaction between magnetic impurity atoms is predicted to decay with the square of the separation, $D$, between the impurities ($D^{-2}$).
  The nature of such an interaction in a graphene system has been the subject of much discussion in recent literature\cite{yazyev_emergence_2010,vozmediano_local_2005,dugaev_exchange_2006,saremi_rkky_2007,brey_diluted_2007,sherafati_rkky_2011,
    black-schaffer_rkky_2010,uchoa_kondo_2011,black-schaffer_importance_2010,power_electronic_2011,sherafati_analytical_2011,power_strain-induced_2012,power_dynamic_2012,venezuela_emergence_2009-1,peng_strain_2012,gorman_rkky_2013}.
  The general consensus from these studies is that the interaction in graphene is shorter ranged due to the vanishing density of states at the Dirac point and decays as $D^{-3}$. 
  Another curiosity is the oscillatory nature of the interaction, which is evident in most materials but hidden in graphene by a commensurability effect between the oscillation periods and the underlying hexagonal lattice.
  Furthermore, the interaction in graphene is predicted to have a sublattice dependence, i.e., magnetic moments on the same sublattice are predicted to align ferromagnetically, whereas those on opposite sublattices will pair antiferromagnetically.
  
  Despite the vast literature regarding the IEC in graphene, the body of work dedicated to edged graphene is surprisingly small\cite{sun_indirect_2013, soriano_hydrogenated_2010, bunder_ruderman-kittel-kasuya-yosida_2009, szalowski_indirect_2013, black-schaffer_rkky_2010, black-schaffer_importance_2010}. 
  This is particularly glaring in light of the fact that many of graphene's local properties (e.g. transport, magnetic and mechanical) are drastically modified near the termination of the graphene lattice \cite{acik_nature_2011, yazyev_emergence_2010, girit_graphene_2009, son_energy_2006, han_energy_2007, power_model_2009, klinovaja_rkky_2013}.
  Thus far, most of the attention has been focused on edges with a zigzag geometry, where localized states were predicted to give rise to spin-polarized edges\cite{fujita_peculiar_1996,nakada_edge_1996,yazyev_emergence_2010,kunstmann_stability_2011}. Such magnetic-edged ribbons have inspired a number of theoretical device proposals\cite{brey_diluted_2007, son_half-metallic_2006, yazyev_magnetic_2008, wang_z-shaped_2007}. Furthermore, success in the accurate patterning of graphene edges \cite{zhang_experimentally_2012,wang_etching_2010,kosynkin_longitudinal_2009} makes this field more accessible from an experimental perspective, and attractive from a theoretical one.
  RKKY interactions mediated by the edge states in zigzag nanoribbons are expected to decay exponentially\cite{black-schaffer_rkky_2010}, in contrast to the power-law decay normally associated with the IEC in metallic systems. However, the presence of electron-electron interactions has been predicted to lead to a distance independent interaction \cite{black-schaffer_importance_2010}. 
  Results in armchair nanoribbons are scarce, however, one study finds an interaction dependent on the distance from the edge\cite{szalowski_indirect_2013} including both an exponential decay and a distance-independent interaction. To our knowledge, the IEC in graphene with a single edge has not been investigated previous to the current work.

  Motivated by the growing interest in the properties of edged graphene, we examine how the IEC between magnetic objects in graphene is affected by their proximity to an edge.  
  Two situations are worth considering, namely the case of a single edge represented by a semi-infinite graphene sheet and the case of two parallel edges which is present in a graphene nanoribbon. In both these situations we will examine pristine armchair geometry edges, and focus on the range of the interaction between two localized magnetic moments with a separation $D_A$ parallel to the edge(s), as illustrated schematically in Figure~\ref{Semi-Infinite_System}.   The interaction range is characterized by the rate of decay $D_{A}^{-\alpha}$ of the IEC and we show that $\alpha$ becomes a function of the distance from the edge in both cases.  In other words, rather than displaying a fixed rate of decay as in bulk graphene, the IEC in edged graphene fluctuates between a long- and short-ranged interaction depending on how far 
  the magnetic objects are from the armchair edges.
  Numerical calculations of the RKKY interaction are backed up by simple analytical expressions which are derived to explain the variations of the interaction range as the proximity to the graphene edges varies. Furthermore, we demonstrate that these variations are the result of interference effects introduced by the presence of one or more edges. 
  
  In what follows we begin by presenting the model used to calculate the IEC and to describe the system under consideration. 
  Special attention is paid to the single-particle Green functions (GFs) of the system, since they are the key ingredients appearing in the IEC expressions. 
  We demonstrate alternative methods to calculate the GFs for a semi-infinite graphene sheet and examine how they relate to their infinite sheet counterparts.
  The GFs derived are very general quantities, and may be used to calculate a wide range of other physical properties in semi-infinite graphene and armchair nanoribbons.
  Results for both cases are presented, followed by a general discussion of the features observed in the numerical calculations and their connection with the analytical model. 
  We finish with a section addressing our conclusions and an outlook on future work in edged graphene systems. 

\section{Model}
\begin{figure}
  \centering
  \includegraphics[width=0.8\textwidth]{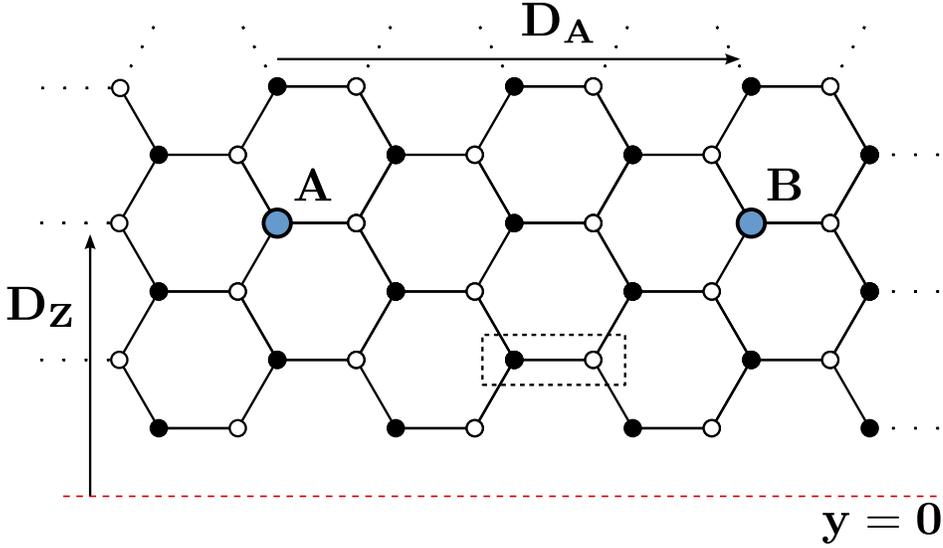}
  \caption{A schematic of our system. 
	    The dotted lines represent a continuation of the system and the dashed red line marks the x-axis (where $D_Z=0$)
	    The dashed rectangle at the center is our two-atom unit cell, containing an atom from each of the two sublattices ($\bullet$ and $\circ$). 
	    The impurities are represented by larger, blue dots. The separation shown is given by $D_Z=4$, $D_A=2$ within our convention.}
  \label{Semi-Infinite_System}
  \end{figure}

  Let us start by defining the system to be studied, consisting of an armchair edged graphene sheet with two embedded magnetic objects, labeled $A$ and $B$, both at a distance $D_Z$ (zigzag direction) from the edge. This setup is shown schematically in Figure~\ref{Semi-Infinite_System}.
  The magnetic objects are assumed to be substitutional magnetic impurities replacing two carbon atoms a distance $D_A$ apart in the armchair direction, although the results are equally valid for top-adsorbed impurities. 
  Note that the distances $D_A$ and $D_Z$ cannot adopt any continuous values but are limited by the constrained integer values defining the hexagonal lattice. 
  In other words, $D_Z$ is calculated in units of $\frac{a}{2}$ (thus counting the number of atoms from the edge), and $D_A$ in units of $\sqrt{3} a$ (thus counting the number of unit cells in the armchair direction), where $a$ is the lattice parameter of graphene. Note that, under this definition, atoms at the very edge of the sheet take the value $D_Z = 1$.

  The IEC between the two magnetic impurities $A$ and $B$ can be calculated as the total energy difference between their ferromagnetic and antiferromagnetic configurations. 
  Using the Lloyd formula method \cite{lloyd_wave_1967, ferreira_fundamental_1996} this energy difference is given by
  \begin{equation}
    J_{BA} = -\frac{1}{\pi} \mathrm{Im} \int_{-\infty}^{\infty} dE \, f(E) \ln (1 + 4 V_{x}^2 G_{BA}^{\uparrow}(E) G_{AB}^{\downarrow}(E) ),
      \label{IEC_equation}
  \end{equation}
  where $V_{x}$ is the exchange splitting between the spin sub-bands of the magnetic impurities, $f(E)$ the Fermi function and
  $G_{BA}^{\sigma}(E)$ is the single-particle GF between the impurities $A$ and $B$ for electrons with spin $\sigma$ with the system in the ferromagnetic configuration. 
  The spin dependent GFs $G^{\sigma}$ can be calculated in terms of their pristine lattice counterparts $g$ 
    by introducing a localized potential term describing the band-splitting at the magnetic impurity sites using Dyson's equation. 
    Within this convention a negative coupling represents favorable ferromagnetic alignment, and a positive coupling an antiferromagnetic one.

  If the coupling is expanded to lowest order in $V_{x}$, we can recover the standard perturbative form of the RKKY interaction, also given by the spin susceptibility of the system
  \begin{equation}
    J_{RKKY} = -\frac{4V_{x}^2}{\pi} \mathrm{Im} \int_{-\infty}^{\infty} dE \, f(E) g_{BA}^2 (E) ,
    \label{RKKY_Equation}
  \end{equation}
  where $g$ is now the GF of the pristine system in the absence of any magnetic impurities.

  To proceed it is important to specify the Hamiltonian defining the electronic structure of graphene, 
    which in this case is described by the nearest-neighbour tight-binding approximation representing the electrons in the $p_z$ orbitals of the carbon atoms:
  \begin{equation}
    \mathcal{H} = \sum_{\mathbf{r},\mathbf{r}'} \sum_{\alpha,\beta} \ket{\mathbf{r},\alpha} t \bra{\mathbf{r'},\beta} \: .
    \label{Hamiltonian}
  \end{equation}
  Graphene is formed from two interlocking triangular sublattices, and our notation is chosen accordingly.
  $\ket{\mathbf{r},\alpha}$ labels a $\pi$ orbital at the site whose unit cell is given by $\mathbf{r}$ and whose sublattice is defined by the index $\alpha = \bullet$ or $\alpha = \circ$ (see Figure~\ref{Semi-Infinite_System}).

  We note from Eqs. (\ref{IEC_equation}) and (\ref{RKKY_Equation}) that the separation dependence of the interaction is entirely within the GFs.
  Thus the distance dependent features of this interaction can be described entirely by examining these quantities, which are derived in the next section.
  
  \subsection{Single-particle Green Functions}
  \label{Sec_SI}
    First we derive the GF for bulk graphene and then show how this approach may be expanded to deal with semi-infinite graphene.
    The GF associated with a Hamiltonian $\mathcal{H}$ is given by
    \begin{equation}
      \hat{g}(E) = \left( E \hat{I} - \hat{\mathcal{H}} \right)^{-1} .
      \label{Green_General}
    \end{equation}
    The Hamiltonian for bulk graphene given in Eq.~(\ref{Hamiltonian}) can be greatly simplified by Fourier transforming to reciprocal space ($\ket{\mathbf{k},\alpha}$), and it may be completely diagonalized by the following choice of eigenstates
    \begin{equation}
      \ket{ \mathbf{k}, \pm} = \frac{1}{\sqrt{2}} \left( \ket{\mathbf{k}, \bullet} \pm \frac{ f^{\dagger}(\mathbf{k}) }{ |f(\mathbf{k})| } \ket{ \mathbf{k}, \circ } \right) ,
    \end{equation}
    where $f (\mathbf{k}) = 1 + 2 e^{i k_x \frac{\sqrt{3}}{2} a} \cos \left( \frac{k_y a}{2} \right)$.
    Diagonalization makes the inversion in Eq.~(\ref{Green_General}) trivial and the GF may now be written as
    \begin{equation}
      \hat{g}(E) = \sum_{\mathbf{k}} \frac{ \ket{\mathbf{k},+} \bra{\mathbf{k},+} }{E - t|f (\mathbf{k})|} + \frac{ \ket{\mathbf{k},-} \bra{\mathbf{k},-} }{E + t|f (\mathbf{k})|}.
      \label{Bulk_Green_General}
    \end{equation}
    This can be projected onto the position basis to get
    \begin{equation}
      \bra{\mathbf{r}_1, \alpha } \hat{g}(E) \ket{\mathbf{r}_2, \beta} = \frac{ a^2 \sqrt{3} }{8 \pi^2} \int dk_y \int dk_x \, \frac{ N^{\alpha \beta}(E,\mathbf{k}) e^{i \mathbf{k} \cdot ( \mathbf{r}_2 - \mathbf{r}_1 ) } }{E^2 - t^2 |f(\mathbf{k})|^2 } ,
    \end{equation}
    where the integration is taken over the first Brillouin Zone in reciprocal space.
    Here $N^{\alpha \beta}(E,\mathbf{k})$ is a sublattice dependent term given by
   \begin{equation}
      N^{\alpha \beta}(E,\mathbf{k}) = \left\{ \begin{array}{ll}
	      E \qquad  & \alpha=\beta \\
	      tf(\mathbf{k}) & \alpha = \bullet; \, \beta = \circ \\
	      tf^{\dagger}(\mathbf{k}) & \alpha = \circ; \, \beta = \bullet \quad .
	      \end{array} \right.
    \end{equation}
    In order to calculate the GF for semi-infinite graphene it is necessary to first obtain a suitable basis of eigenstates.
    We consider the following linear combination of bulk states
    \begin{equation}
      \ket{ \phi_{\mathbf{k}}, \pm } = \frac{1}{\sqrt{2}} \left( \ket{ (k_x , k_y), \pm } - \ket{ (k_x, - k_y), \pm } \right),
      \label{Semi_Eigenstates}
    \end{equation}
    where the vector $\mathbf{k}$ is written explicitly in terms of its components $k_x$ and $k_y$, and we examine its projection on the position basis
    \begin{eqnarray*}
      \braket{(x,y),\bullet}{ \phi_{\mathbf{k}}, \pm } &=& -\frac{i}{\sqrt{N}} e^{-i k_x x} \sin(k_y y) \\
      \braket{(x,y),\circ}{ \phi_{\mathbf{k}}, \pm } &=& \mp \frac{i}{\sqrt{N}} e^{-i k_x x} \sin(k_y y) \frac{ f^{\dagger}(\mathbf{k}) }{ |f(\mathbf{k})| }.
    \end{eqnarray*}
    It is clear that the projection vanishes whenever $y=0$, indicating that there are no longer any atomic states along the x-axis.
    This means that the graphene sheet has been divided up into two equivalent halves, the layout of which can be seen in Fig.~\ref{Image_Geometry}.
    
    With the new choice of eigenstates the semi-infinite GF is given by
    \begin{equation}
      \hat{S}(E) = \sum \frac{ \ket{\phi_{\mathbf{k}}, + } \bra{\phi_{\mathbf{k}}, + } }{E - \epsilon_+} + \frac{ \ket{\phi_{\mathbf{k}}, - } \bra{\phi_{\mathbf{k}}, - } }{E + \epsilon_+} ,
	\label{Semi_Green_general}
    \end{equation}
    and this can be similarly projected onto the position basis to get
    \begin{eqnarray}
      \bra{(x_1,y_1)  ,\alpha}  S(E) && \ket{(x_2,y_2),\beta} = \frac{a^2 \sqrt{3}}{4\pi^2} \int dk_y \int dk_x \, \nonumber \\
	 &&\times \frac{ N^{\alpha \beta}(E,\mathbf{k}) e^{i k_x (x_2 - x_1)} \sin(k_y y_1) \sin(k_y y_2) }{E^2 - t^2 |f(\mathbf{k})|^2} .
    \end{eqnarray}
    One of these integrals can be solved by contour integration, upon which the GF between two atoms a distance $D_Z$ from the edge and separated by a distance $D_A$ in the armchair direction is given by
    \begin{equation}
      S(E,D_Z,D_A) = \frac{i}{2\pi t^2} \int_{-\frac{\pi}{2}}^{\frac{\pi}{2}} dk_Z \, \frac{ E \, e^{2iqD_A} \sin^2(k_Z D_Z) }{\cos(k_Z)\sin(q)} ,
	\label{Semi_Green_single}
    \end{equation}
    where $q$ is the pole from the contour integration and is given by
    \begin{equation}
      q = \pm \cos^{-1} \left[ \frac{E^2 - t^2 - 4t^2 \cos^2 (k_Z)}{4t^2 \cos(k_Z)} \right] ,
    \end{equation}
    and the sign of the pole is chosen such that its imaginary part is always positive.
    Here we have introduced the dimensionless k-space vectors $k_A=\frac{\sqrt{3}a}{2} k_x$ and $k_Z = \frac{a}{2} k_y$.
    Numerical integration gives the GF exactly, but we can also use the stationary phase approximation (SPA) to find an approximate analytic form of the GF, and we do so later in this paper.

    The fact that we have formed the semi-infinite GF by considering combinations of bulk eigenstates makes it natural to see whether the semi-infinite GF can be written in terms of its bulk counterparts.
    It turns out that this is both possible to do, and provides geometric insight into its meaning.

  \subsection{The Image Method}
    As previously mentioned, the IEC in bulk graphene has been extensively studied and most of its features are well understood.
    Therefore, our strategy is to express the semi-infinite system representing a single-edged graphene sheet in terms of its bulk counterpart so that we can infer the IEC behavior in the presence of edges.
    Such a strategy requires that we relate the semi-infinite GF to that of bulk graphene.

\begin{figure}
      \centering
      \includegraphics[width=0.8\textwidth]{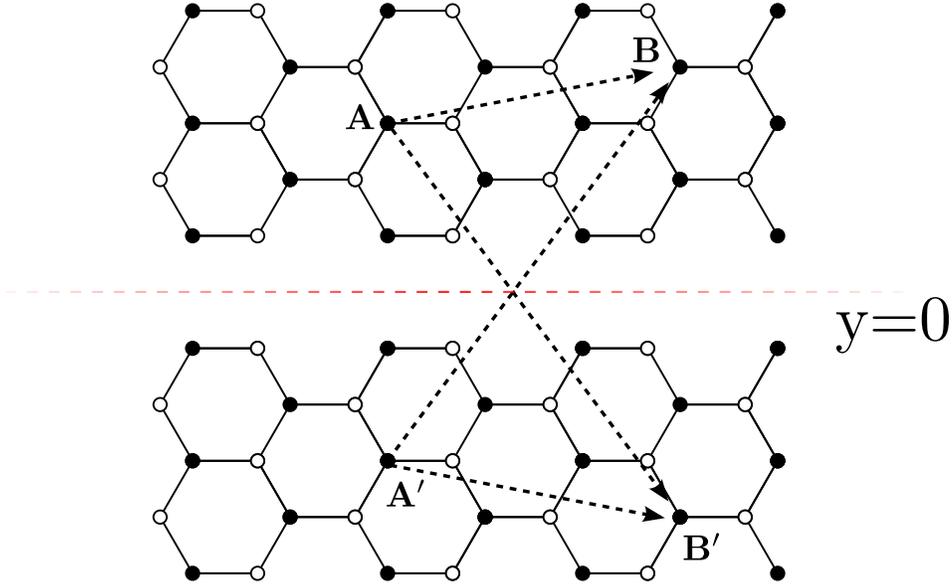}
      \caption{A semi-infinite Green function between $A$ and $B$ can be written as a sum of bulk terms between $A$, $B$ and their image sites $A'$, $B'$}
      \label{Image_Geometry}
    \end{figure}

    In order to express the semi-infinite GF in terms of the more familiar bulk GFs we expand out the eigenstates in Eq.~(\ref{Semi_Green_general}) in terms of their bulk equivalents of Eq.~(\ref{Semi_Eigenstates}).
    Collecting all the positive $k_y$ terms $(\ket{(k_x,k_y),\pm})$ leads to the general formula for the bulk GF (Eq.~\ref{Bulk_Green_General}). 
    The problem that arises is in dealing with terms that have a $-k_y$, such as
    \begin{equation}
      \hat{g}_2(E) = \sum \frac{ \ket{(k_x,k_y),+} \bra{(k_x,-k_y),+} }{E - \epsilon_+} 
      + \frac{ \ket{(k_x,k_y),-} \bra{(k_x,-k_y),-} }{E + \epsilon_+} .
    \end{equation}
    These quantities can be dealt with by noting that, when dealing with the projection of a reciprocal-space state onto the real space, the minus sign can be freely swapped between $k_y$ and $y$.
    By means of a Fourier transform, it can be shown that
    \begin{equation}
      \bra{ (x,y), \alpha} (k_x, -k_y), \pm \rangle = \bra{ (x,-y) } (k_x, k_y) , \pm \rangle .
    \end{equation}
    Thus a sign change in $k_y$ is equivalent to a reflection of the position vector about the y-axis.
    Now $\bra{\mathbf{r}_1,\alpha} g_2 (E) \ket{ \mathbf{r}_2 , \beta}$ can be seen as the GF between $\mathbf{r}_1$ and the reflection of $\mathbf{r}_2$.
    This method can be applied to the remaining factors, allowing us to write the semi-infinite GF between $A$ and $B$ as 
    \begin{eqnarray}
      S_{AB}(E) &=& \frac{1}{2} \big( g_{AB}(E) - g_{AB'}(E) - g_{A'B}(E) + g_{A'B'}(E) \big) \nonumber \\
      &=& g_{AB}(E) - g_{AB'}(E) \,,
      \label{SGF_fn_of_GF}
    \end{eqnarray}
    where $A'$ and $B'$ are the images of $A$ and $B$ as shown in Fig.~\ref{Image_Geometry}, which also underlines the simplicity and intuitive nature of the approach. 
    From Eq. \ref{SGF_fn_of_GF} and Fig.~\ref{Image_Geometry} we see that the Green function connecting two sites on the semi-infinite lattice is equivalent to that on the infinite lattice with the addition of a correction term that takes into account the edge-induced scattering. 
    This correction term is simply the Green function connecting one site with the site equivalent to the image of the other in the infinite graphene sheet with a phase shift of $\pi$. 
    This rather intuitive and computationally convenient result is a direct consequence of the simple manner in which the bands of armchair-edged graphene can be written in terms of their infinite sheet counterparts. 
    For other edge geometries, the presence of localized edge states which cannot easily be reconciled with the infinite graphene band structure complicate the picture and prevent the 
      formulation of the simple relation given in  Eq. \ref{SGF_fn_of_GF}. Furthermore the entire $D_Z$ dependence of the semi-infinite sheet GF is contained within this correction term, and from the standard behavior of the graphene GF we can determine that it will decay with increased distance from the edge as $\frac{1}{\sqrt{D_Z}}$. Thus, when dealing with atoms a long distance from the edge, we will find that $g_{AB'}(E) << g_{AB}(E)$, and the bulk GFs are recovered.

    In addition to the expressions presented above, GFs for both the nanoribbon and semi-infinite graphene can be calculated numerically via recursive techniques whereby Dyson's equation is used to join ``strips'' of graphene.
    All of the calculations in this paper were also performed recursively, within the Rubio-Sancho scheme\cite{sancho_quick_1984}, and were found to be in perfect agreement with the integration methods presented here.
    Given numerical expressions for the GFs, it becomes possible to explore the interaction computationally.

\section{RKKY Interaction in Semi-Infinite Graphene}
  \subsection{Numerical Results}
\begin{figure}
      \centering
      \includegraphics[width=0.8\textwidth]{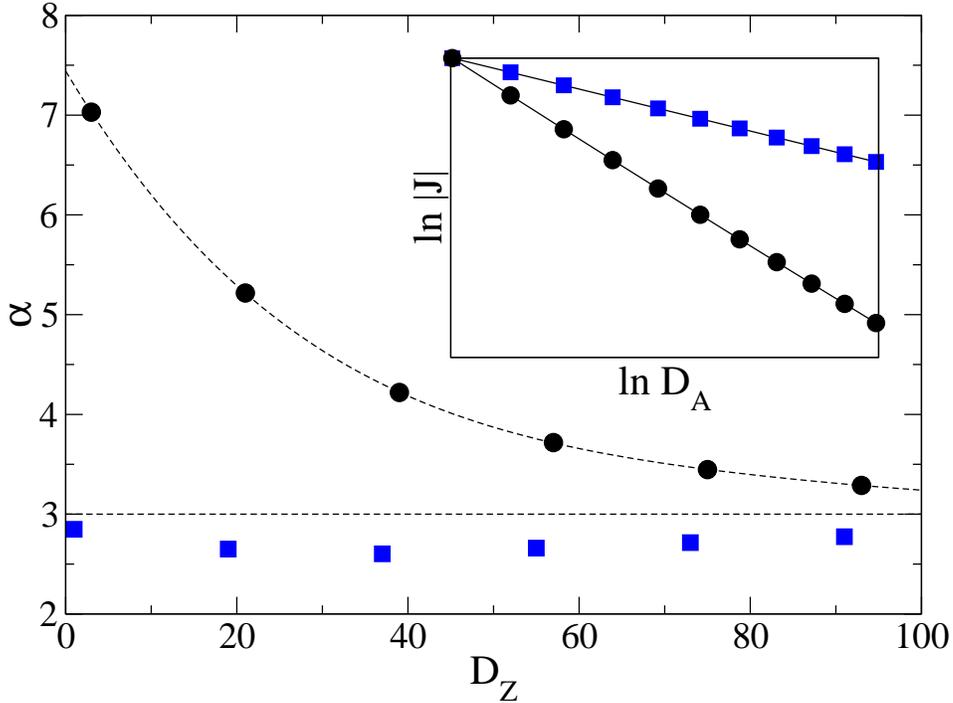}
      \caption{A plot of the decay exponent $\alpha$ against $D_Z$ for various representative values.
		  We see that the slower rate of decay (blue squares) remains close to the bulk value of $\alpha=3$ at all times.
		  The more rapid decay (black circles) starts from a value of $D_Z=7$ but has become almost identical to the bulk value once we reach $D_Z=100$.
		  Inset is a log-log plot of the coupling($J$) against $D_A$ for $D_Z=1$ (blue squares) and $D_Z=3$ (black circles). 
		  The rate of the decay is given by the slope, and is significantly faster for the $D_Z = 3$ case.}
      \label{DecayComparison}
    \end{figure}

    Using Eq.~(\ref{IEC_equation}), the IEC was calculated for a number of different value of $D_Z$, the distance from the edge. 
    For each $D_Z$ value, the separation between the impurities, $D_A$, was increased and this data was used to determine the power-law decay rate. 
    In all cases a power-law expression of the form $J_{BA} \sim D_A^{-\alpha}$ fitted the data to very high accuracy as evinced by the log-log plot in the inset of Fig.~\ref{DecayComparison} for $D_Z=1$ (blue squares) and  $D_Z=3$ (black circles).
    The decay exponent, $\alpha$, is plotted for a number of representative $D_Z$ values in the main panel of Fig.~\ref{DecayComparison}.
    We note that in most cases (blue squares) the rate of decay was identical to that of bulk ($\alpha = 3$).
    However whenever $D_Z$ was a multiple of 3 (black circles), much more rapid decay rates were recovered, starting at more than $D^{-7}$ for atoms near the edge.
    Further investigation showed that this short-ranged decay converged asymptotically to its longer-ranged counterpart as the impurities were moved further from the edge into the graphene sheet. This behavior is illustrated by the dashed lines in Fig.~\ref{DecayComparison} showing the fixed $D^{-3}$ decay and the variable, faster ranged decay as a function of $D_Z$.
    
    These calculations were repeated for sites on different sublattices, where we found the same qualitative features with the opposite sign, indicating an antiferromagnetic alignment. It is natural to try and interpret these results analytically, and we explain our method for doing so in the next section.

  \subsection{Analytical Results}
\begin{figure}
      \centering
      \includegraphics[width=0.8\textwidth]{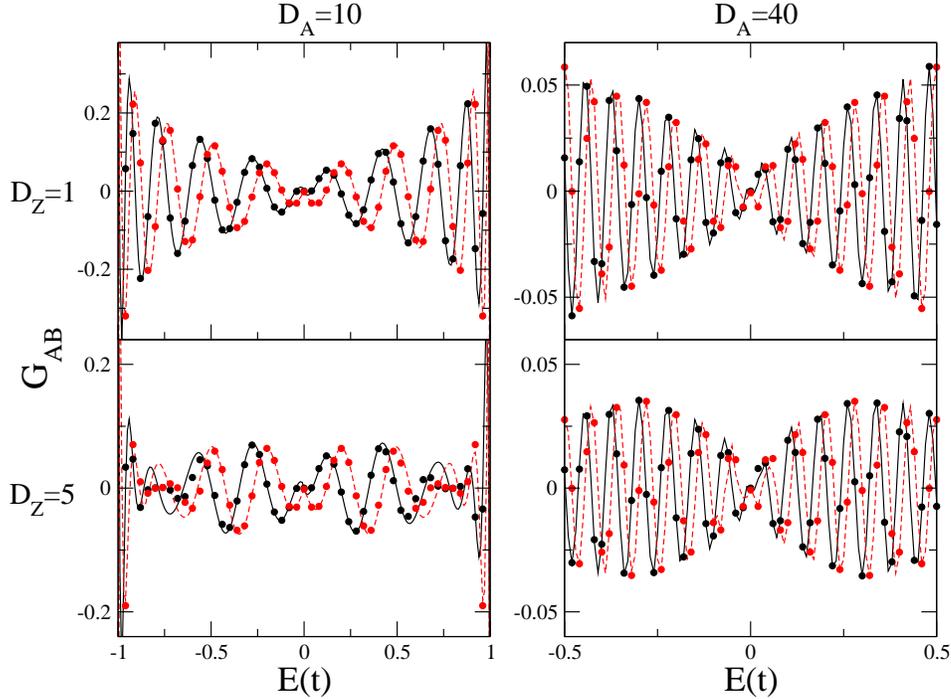}
      \caption{A comparison of numerical integration against the SPA for various armchair separations/edge separations.
	The black lines represent the real part of the GF and the red lines the imaginary part.
	The corresponding dots represent the SPA approximation.
	The top two graphs are performed at $D_Z = 1$ and the bottom two at $D_Z = 5$.
	Those on the left are performed for $D_A = 10$ and those on the right for $D_A = 40$.
	It is clear that best agreement occurs when close to the edge and at large separations.}
      \label{SPA_Comp}
    \end{figure}

    For large separations, the bulk GF has been shown to be well approximated over the whole energy band by the SPA, particularly when working in the armchair or zigzag directions\cite{power_electronic_2011}.
    The SPA takes advantage of the highly oscillatory nature of the Green functions, demonstrating that, due to destructive interference, an excellent approximation can be attained by considering only a small part of the integrand.
    The same approach can be used to solve the remaining integral in Eq.~(\ref{Semi_Green_single}), giving us
    \begin{equation}
      S(E) = \frac{ \mathcal{A}(E,D_Z) e^{i Q(E) D_A} }{\sqrt{D_A}} ,
      \label{Green_SPA}
    \end{equation}
    where
    \begin{equation}
      \mathcal{A}(E,D_Z) = \pm 2iE \,
	\frac{ \sin^2 \left[ D_Z \sin^{-1} \left( \frac{ \sqrt{E^2+3t^2} }{2t} \right) \right] }
	{ \sqrt{E^2+3t^2} \left[E^2(t^2-E^2)\right]^{\frac{1}{4}} } \sqrt{ \pm \frac{i}{\pi} } ,
    \end{equation}
    and
    \begin{equation}
      Q(E) = \pm 2 \cos^{-1} \left( \frac{ \sqrt{t^2-E^2} }{t} \right) ,
    \end{equation}
    and $\pm = -\mathrm{Sign}[E]$.
    The SPA provides an excellent approximation to the exact numerical integration, which can be seen in Fig.~\ref{SPA_Comp}.
    When working within the SPA, best results are obtained when the exponential part of the integrand oscillates much faster than the rest.
    As seen from Eq.~(\ref{Semi_Green_single}) this implies that our approximation will be most accurate close to the edge and for large separations between the magnetic objects (e.g. when $D_A \gg D_Z$).
    
    Putting our GF in the form given in Eq.~\ref{Green_SPA} simplifies the analytic calculation of the coupling immensely. 
    Substituting this into the formula for the RKKY coupling (Eq.~\ref{RKKY_Equation}) gives
    \begin{equation}
      J_{BA}^{RKKY} = \frac{V_{ex}^2}{D_A} \, \mathrm{Im} \int dE \, \frac{ \mathcal{B}(E) e^{2iQ(E) D_A} }{1 + e^{\beta(E-E_F)}} ,
    \end{equation}
    where $\mathcal{B}(E,D_Z) = \mathcal{A}^2 (E,D_Z)$.
    This equation can be solved via contour integration in the upper half plane and a summation over the Matsubara frequencies\cite{kirwan_sudden_2008}.
    Finally, expanding around the Fermi energy and taking the low temperature limit gives
    \begin{equation}
    J^{RKKY}_{BA} = -V_{ex}^2 \, \mathrm{Im} \sum_{\ell=0}^{\infty} \frac{ \mathcal{B}^{(\ell)} (E) e^{2iQ^{(0)}(E) D_A} }{ (2iQ^{(1)}(E))^{\ell+1} D_A^{\ell+2} } ,
    \end{equation}
    where a superscript of $(\ell)$ denotes the $\ell^{th}$ derivative evaluated at the Fermi energy.
    This equation implies that the decay rate of the coupling is determined entirely by the first non-vanishing $\mathcal{B}^{(\ell)}(E)$.
    $\mathcal{B}(E)$ always vanishes when evaluated at $E_F=0$. 
    However, its first derivative is given by 
    \begin{equation}
      \mathcal{B}^{(1)}(E) = \frac{4 i \sin^4 \left[\frac{D_Z \pi }{3}\right]}{3 \pi t^3},
    \end{equation}
    which only vanishes when $D_Z$ is a multiple of 3. Higher order derivatives are also found to vanish.
    The analytical expressions thus predict both a long-ranged $D_A^{-3}$ interaction and the presence of a shorter-ranged interaction at every third distance from the edge. 
    This is in agreement with the numerical results discussed earlier and suggests that RKKY interactions are likely to be highly influenced by the presence of a nearby edge. 
    Although these calculations have been performed with both impurities the same distance from the edge, we have also performed similar calculations for a wider range of cases and find that when either impurity is a multiple of 3 from the edge a faster decay rate arises.

    The same approach that we have developed to deal with semi-infinite graphene can be extended to study the case of the coupling between magnetic objects in graphene nanoribbons.

\section{Results in Metallic armchair nanoribbons}
  Although some previous works have examined RKKY interactions in nanoribbons, there is some discrepancy between these as to the exact behavior. 
  One work\cite{black-schaffer_rkky_2010} suggests that, away from zigzag edges, little difference will be seen from the bulk graphene result, whereas another\cite{szalowski_indirect_2013} predicts more unusual behavior in armchair nanoribbons, including a potentially separation-independent (i.e. non-decaying) interaction. 
  The difficulty in patterning atomically precise nanoribbon devices may hinder the spintronic application of the IEC in such systems, however the potential for longer-ranged interactions stemming from the lower system dimensionality motivates further examination of nanoribbon-based IEC. 
  Furthermore, we expect that the behavior in nanoribbons will be strongly connected to that already discussed for semi-infinite graphene and that parallels may be drawn between the two systems.
  We consider only metallic armchair nanoribbons (of width $N=3n+2$, where $N$ is the number of atoms in a zigzag strip across the nanoribbon and $n$ is an integer), since the IEC decays exponentially fast in semi-conducting nanoribbons\cite{szalowski_indirect_2013}.
  
  The GFs for metallic armchair nanoribbons can be determined from those in semi-infinite graphene simply by introducing a suitable quantization condition that forces the projection of the wavefunctions to vanish on both sides of the nanoribbon.
  Requiring that the projection vanishes at $0$ and $n_E \frac{a}{2}$ defines an armchair nanoribbon of width $N = n_E -1$ and quantizes the wavevector in the zigzag direction so that we have $k_Z = \frac{\pi j}{n_E}$ where $j = 0 \rightarrow n_E - 1$.
  Then following the same integration methods as Section~\ref{Sec_SI} we obtain an expression for the Green function between two sites equidistant from the edge in an armchair nanoribbon
  \footnote{Special care must be taken when dealing with terms where $j=\frac{n_E}{2}$ due to the singularity in the denominator. 
    By examining the limiting behavior of this term carefully, it can be shown that it does not contribute to the GF for finite $D_A$, and contributes in the form of a modified expression when $D_A=0$.}
  \begin{equation}
    \hat{S}_{GNR}(E) = \frac{i}{2 n_E t^2} \sum_{j} \frac{ N^{\alpha \beta}e^{2 i q D_A} \sin^2 \left[\frac{\pi j}{n_E} D_Z \right] }{ \cos(\frac{\pi j}{n_E}) \sin(q) },
    \label{GNR_GF}
  \end{equation}
  and
  \begin{equation}
    q_{GNR} = \pm \cos^{-1} \left[ \frac{E^2-t^2-4 t^2 \cos^2 (\frac{\pi j}{n_E})}{ 4t^2 \cos( \frac{\pi j}{n_E} ) } \right].
  \end{equation}
  Note that here $D_Z$ refers to the distance of both sites from one of the edges of the nanoribbon.
  As with our semi-infinite results, the calculations were repeated using recursive techniques for comparison and yielded identical results.
  
\begin{figure}
  \centering
  \includegraphics[width=0.8\textwidth]{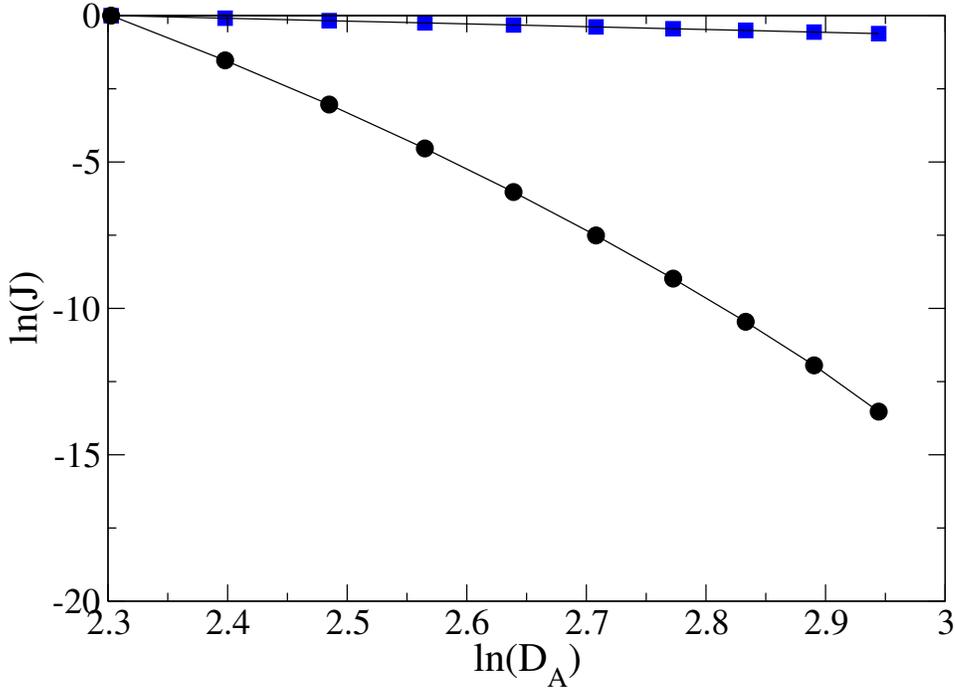}
  \caption{A log-log plot of the coupling $J$ against armchair distance $D_A$.
      The blue squares represent $D_Z = 1$ and the black circles represent $D_Z = 3$.
      The former case corresponds to an interaction decaying as $D_A^{-1}$ whereas the latter decays exponentially. 
      These calculations were performed for a ribbon width of $n_E=6$}
  \label{Ribbon_Coupling}
  \end{figure}

  The coupling can be determined numerically by using Eq.~\ref{IEC_equation} and the recently developed GFs. 
  Some sample results for the coupling are plotted in Fig.~\ref{Ribbon_Coupling}, from which we may draw an immediate similarity with the inset in Fig.~\ref{DecayComparison}. For different values of $D_Z$ we note distinctly different types of interaction.
  In contrast with the semi-infinite case, here we find that the `slow' decay is more gentle and the rapid decay does not obey a power law behavior.
  We find that, for most values of $D_Z$, the coupling decays as $D^{-1}$, which matches some predictions for armchair nanoribbons\cite{szalowski_indirect_2013}, nanotubes\cite{costa_indirect_2005} and is what would be expected in a 1d electron gas\cite{yafet_ruderman-kittel-kasuya-yosida_1987}.
  However, when $D_Z$ is a multiple of 3, we find that our plots are better fitted with an exponential decay.
  This behavior has been verified for a wide range of different $D_Z$ and ribbons widths.
  We note that this behavior is similar to that reported previously in the literature\cite{szalowski_indirect_2013}, however we do not observe the non-decaying interaction predicted for certain cases in that work.

\begin{figure}
  \centering
  \includegraphics[width=0.8\textwidth]{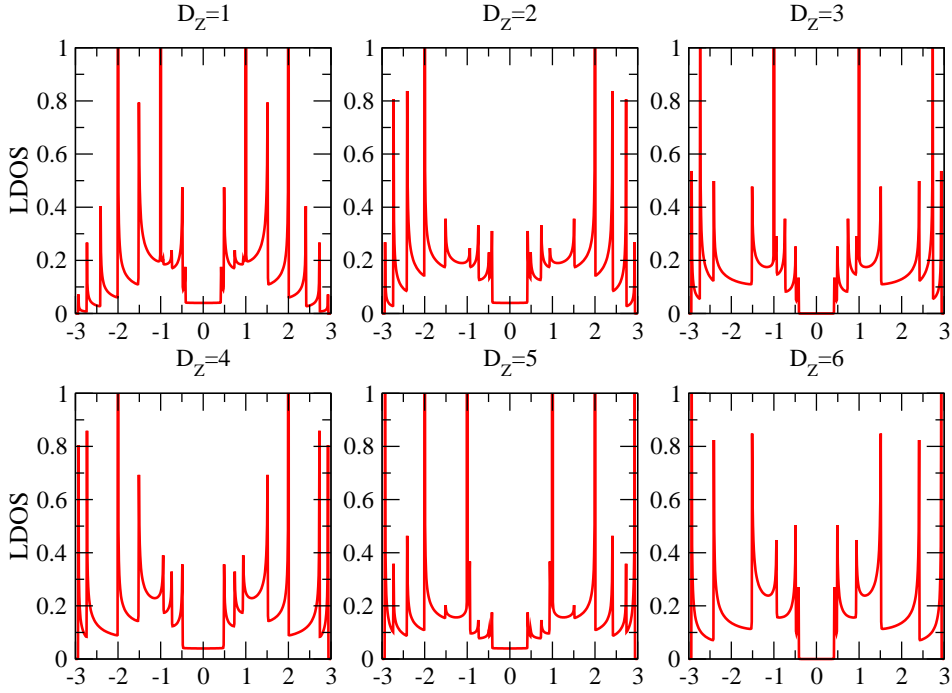}
  \caption{A plot of the local density of states for various distances from the edge ($D_Z$). The LDOS disappears at the Fermi energy when $D_Z$ is a multiple of 3. This leads to the exponential decay of the IEC at these $D_Z$ values.}
  \label{RibbonDOS}
  \end{figure}

  The fact that we find an exponential decay for certain $D_Z$ values is not surprising when we examine the electronic structure of metallic armchair nanoribbons. 
  It is known that the LDOS in nanoribbons vanishes at the half-filled Fermi energy for atoms a multiple of 3 from the edge. 
  The LDOS is given by the imaginary part of the diagonal GF, and it can be shown by examining Eq. (\ref{GNR_GF}) at $D_A=0$ and $E=0$ that this is zero whenever $D_Z$ is a multiple of 3. 
  This behavior is shown in detail in Figure~\ref{RibbonDOS}.
  The vanishing DOS means that there are no conducting states around the Fermi energy and thus the interaction decays exponentially for these $D_Z$ values, as it would in a semiconductor.\cite{costa_indirect_2005}.

\section{Conclusions}
  In summary, we studied the IEC in armchair-edged graphene using a variety of different techniques for calculating the necessary Green functions.
  The Green functions methods derived allow for quick calculation of many properties of semi-infinite graphene, and allow for easy comparison to bulk graphene cases due to their simple relation to the GFs for an infinite graphene sheet. 
  An analytic approximation for armchair separations in such systems using the SPA method yielded simple closed form expressions in excellent agreement with more intensive numerical calculations for a wide range of cases.
  The fully analytical GF derived for armchair nanoribbons should similarly prove useful for more efficient analytic and numerical investigation of these systems.
  In both the semi-infinite and nanoribbon systems a three-fold pattern of unusually rapid decay rates for the IEC was identified. 
  Such behavior is strongly dependent on the distance from the armchair edge and is not present in bulk graphene systems.
  In nanoribbons, this faster decay rate was found to depend exponentially on the separation between magnetic moments, whereas in semi-infinite graphene it was determined to follow a power-law behavior.
  The longer ranged interaction shown here for nanoribbons has potential application in the field of spintronics, and the ability to further tune the interaction by exploiting the threefold behavior cannot be ruled out.
  However, the main importance of this result is that it completes the spectrum of behavior for different armchair-edged graphene allotropes.
  We see that the triplet exponential behavior displayed by nanoribbons softens to a power law decay in single-edged graphene, and eventually converges to the bulk decay rate of $D^{-3}$ everywhere when the magnetic objects are located far enough from the edge.
  Thus we have identified a clear behavioral pattern taking graphene from the double-edged to the edgeless cases.
  Our results outline a possible avenue for tuning interaction strength by controlling impurity adsorption sites in both armchair-edge graphene and armchair-edge GNRs.
  
  The success we have encountered here encourages the possibility of applying our methodology to different, but similar, systems. 
  The simplest extension of our work would be in examining different edge geometries, in particular either a zigzag or mixed geometry.
   The behavior of multiple impurities is also a topic of much importance, and it is worth investigating if the overall alignment of randomly distributed moments will vary near an edge due to the effects noted here.
  Finally, we would like to investigate ways of extending the interaction range, since a long-ranged, strong interaction is paramount for device applications.
  One well known way of extending the range of the interaction between magnetic moments is to set one of the moments to precess\cite{heinrich_dynamic_2003,guimaraes_spin_2011}.
  Indeed, theoretical studies in graphene have already predicted a large increase in the interaction range in this situation\cite{power_dynamic_2012}.
  A natural extension of this work is to examine this same dynamic interaction in the presence of edges, determining whether or not the threefold pattern of rapid decay remains.
  In conclusion, the results presented here offer a theoretical framework for future studies of magnetic interactions in edged graphene.
  
\section*{Acknowledgments}
  The authors acknowledge financial support received from the Programme for Research in Third-Level Institutions PRTLI5 Ireland, the Irish Research Council for Science, Engineering and Technology under the EMBARK initiative and from Science Foundation Ireland under Grant No. SFI 11/RFP.1/MTR/3083. 
  The Center for Nanostructured Graphene (CNG) is sponsored by the Danish National Research Foundation, Project No. DNRF58.


\bibliography{biblio}

\begin{thebibliography}{10}

\bibitem{geim_rise_2007}
A.~K. Geim and K.~S. Novoselov.
\newblock The rise of graphene.
\newblock {\em Nature Materials}, 6(3):183--191, 2007.

\bibitem{geim_graphene:_2009}
A.~K. Geim.
\newblock Graphene: Status and prospects.
\newblock {\em Science}, 324(5934):1530--1534, June 2009.

\bibitem{pesin_spintronics_2012}
Dmytro Pesin and Allan~H. {MacDonald}.
\newblock Spintronics and pseudospintronics in graphene and topological
  insulators.
\newblock {\em Nature Materials}, 11(5):409--416, 2012.

\bibitem{kane_quantum_2005}
C.~L. Kane and E.~J. Mele.
\newblock Quantum spin hall effect in graphene.
\newblock {\em Physical Review Letters}, 95(22):226801, November 2005.

\bibitem{balandin_superior_2008}
Alexander~A. Balandin, Suchismita Ghosh, Wenzhong Bao, Irene Calizo, Desalegne
  Teweldebrhan, Feng Miao, and Chun~Ning Lau.
\newblock Superior thermal conductivity of single-layer graphene.
\newblock {\em Nano Letters}, 8(3):902--907, March 2008.

\bibitem{huertas-hernando_spin-orbit_2006}
Daniel Huertas-Hernando, F.~Guinea, and Arne Brataas.
\newblock Spin-orbit coupling in curved graphene, fullerenes, nanotubes, and
  nanotube caps.
\newblock {\em Physical Review B}, 74(15):155426, October 2006.

\bibitem{yazyev_hyperfine_2008}
Oleg~V. Yazyev.
\newblock Hyperfine interactions in graphene and related carbon nanostructures.
\newblock {\em Nano Letters}, 8(4):1011--1015, April 2008.

\bibitem{ruderman_indirect_1954}
M.~A. Ruderman and C.~Kittel.
\newblock Indirect exchange coupling of nuclear magnetic moments by conduction
  electrons.
\newblock {\em Physical Review}, 96(1):99--102, October 1954.

\bibitem{yafet_ruderman-kittel-kasuya-yosida_1987}
Y.~Yafet.
\newblock Ruderman-kittel-kasuya-yosida range function of a one-dimensional
  free-electron gas.
\newblock {\em Physical Review B}, 36(7):3948--3949, September 1987.

\bibitem{edwards_oscillations_1991}
D.~M. Edwards, J.~Mathon, R.~B. Muniz, and M.~S. Phan.
\newblock Oscillations in the exchange coupling of ferromagnetic layers
  separated by a nonmagnetic metallic layer.
\newblock {\em Journal of Physics: Condensed Matter}, 3(26):4941, July 1991.

\bibitem{dalbuquerque_e_castro_theory_1994}
J.~{d'Albuquerque}~e Castro, M.~S. Ferreira, and R.~B. Muniz.
\newblock Theory of the exchange coupling in magnetic metallic multilayers.
\newblock {\em Physical Review B}, 49(22):16062--16065, June 1994.

\bibitem{ferreira_nature_1994}
M~S Ferreira, R~B Muniz, J~{d'Alberquerque}~e Castro, and D~M Edwards.
\newblock The nature and validity of the {RKKY} limit of exchange coupling in
  magnetic trilayers.
\newblock {\em Journal of Physics: Condensed Matter}, 6(40):L619--L623, October
  1994.

\bibitem{baibich_giant_1988}
M.~N. Baibich, J.~M. Broto, A.~Fert, F.~Nguyen Van~Dau, F.~Petroff, P.~Etienne,
  G.~Creuzet, A.~Friederich, and J.~Chazelas.
\newblock Giant magnetoresistance of ({001)Fe/(001)Cr} magnetic superlattices.
\newblock {\em Physical Review Letters}, 61(21):2472--2475, November 1988.

\bibitem{binasch_enhanced_1989}
G.~Binasch, P.~Gr\"{u}nberg, F.~Saurenbach, and W.~Zinn.
\newblock Enhanced magnetoresistance in layered magnetic structures with
  antiferromagnetic interlayer exchange.
\newblock {\em Physical Review B}, 39(7):4828--4830, March 1989.

\bibitem{bruno_oscillatory_1991}
P.~Bruno and C.~Chappert.
\newblock Oscillatory coupling between ferromagnetic layers separated by a
  nonmagnetic metal spacer.
\newblock {\em Physical Review Letters}, 67(12):1602--1605, September 1991.

\bibitem{kaiser_magnetic_2007}
Uwe Kaiser, Alexander Schwarz, and Roland Wiesendanger.
\newblock Magnetic exchange force microscopy with atomic resolution.
\newblock {\em Nature}, 446(7135):522--525, March 2007.

\bibitem{meier_revealing_2008}
Focko Meier, Lihui Zhou, Jens Wiebe, and Roland Wiesendanger.
\newblock Revealing magnetic interactions from single-atom magnetization
  curves.
\newblock {\em Science}, 320(5872):82--86, April 2008.
\newblock {PMID:} 18388289.

\bibitem{zhou_strength_2010}
Lihui Zhou, Jens Wiebe, Samir Lounis, Elena Vedmedenko, Focko Meier, Stefan
  Bl\"{u}gel, Peter~H. Dederichs, and Roland Wiesendanger.
\newblock Strength and directionality of surface {Ruderman Kittel Kasuya
  Yosida} interaction mapped on the atomic scale.
\newblock {\em Nature Physics}, 6(3):187--191, March 2010.

\bibitem{yazyev_emergence_2010}
Oleg~V Yazyev.
\newblock Emergence of magnetism in graphene materials and nanostructures.
\newblock {\em Reports on Progress in Physics}, 73(5):056501, May 2010.

\bibitem{vozmediano_local_2005}
M.~A.~H. Vozmediano, M.~P. L\'{o}pez-Sancho, T.~Stauber, and F.~Guinea.
\newblock Local defects and ferromagnetism in graphene layers.
\newblock {\em Physical Review B}, 72(15):155121, October 2005.

\bibitem{dugaev_exchange_2006}
V.~K. Dugaev, V.~I. Litvinov, and J.~Barnas.
\newblock Exchange interaction of magnetic impurities in graphene.
\newblock {\em Physical Review B}, 74(22):224438, December 2006.

\bibitem{saremi_rkky_2007}
Saeed Saremi.
\newblock {RKKY} in half-filled bipartite lattices: Graphene as an example.
\newblock {\em Physical Review B}, 76(18):184430, November 2007.

\bibitem{brey_diluted_2007}
L.~Brey, H.~A. Fertig, and S.~Das~Sarma.
\newblock Diluted graphene antiferromagnet.
\newblock {\em Physical Review Letters}, 99(11):116802, September 2007.

\bibitem{sherafati_rkky_2011}
M.~Sherafati and S.~Satpathy.
\newblock {RKKY} interaction in graphene from the lattice green's function.
\newblock {\em Physical Review B}, 83(16):165425, April 2011.

\bibitem{black-schaffer_rkky_2010}
Annica~M. Black-Schaffer.
\newblock {RKKY} coupling in graphene.
\newblock {\em Physical Review B}, 81(20):205416, May 2010.

\bibitem{uchoa_kondo_2011}
Bruno Uchoa, T.~G. Rappoport, and A.~H. Castro~Neto.
\newblock Kondo quantum criticality of magnetic adatoms in graphene.
\newblock {\em Physical Review Letters}, 106(1):016801, January 2011.

\bibitem{black-schaffer_importance_2010}
Annica~M. Black-Schaffer.
\newblock Importance of electron-electron interactions in the {RKKY} coupling
  in graphene.
\newblock {\em Physical Review B}, 82(7):073409, August 2010.

\bibitem{power_electronic_2011}
S.~R. Power and M.~S. Ferreira.
\newblock Electronic structure of graphene beyond the linear dispersion regime.
\newblock {\em Physical Review B}, 83(15):155432, April 2011.

\bibitem{sherafati_analytical_2011}
M.~Sherafati and S.~Satpathy.
\newblock Analytical expression for the {RKKY} interaction in doped graphene.
\newblock {\em Physical Review B}, 84(12):125416, September 2011.

\bibitem{power_strain-induced_2012}
S.~R. Power, P.~D. Gorman, J.~M. Duffy, and M.~S. Ferreira.
\newblock Strain-induced modulation of magnetic interactions in graphene.
\newblock {\em Physical Review B}, 86(19):195423, November 2012.

\bibitem{power_dynamic_2012}
S.~R. Power, F.~S.~M. Guimar\~{a}es, A.~T. Costa, R.~B. Muniz, and M.~S.
  Ferreira.
\newblock Dynamic {RKKY} interaction in graphene.
\newblock {\em Physical Review B}, 85(19):195411, May 2012.

\bibitem{venezuela_emergence_2009-1}
P.~Venezuela, R.~B. Muniz, A.~T. Costa, D.~M. Edwards, S.~R. Power, and M.~S.
  Ferreira.
\newblock Emergence of local magnetic moments in doped graphene-related
  materials.
\newblock {\em Physical Review B}, 80(24):241413, December 2009.

\bibitem{peng_strain_2012}
Feng Peng and Wei Hongbin.
\newblock Strain enhanced exchange interaction between impurities in graphene.
\newblock {\em Physica B: Condensed Matter}, 407(17):3434--3436, September
  2012.

\bibitem{gorman_rkky_2013}
P.~D. Gorman, J.~M. Duffy, M.~S. Ferreira, and S.~R. Power.
\newblock {RKKY} interaction between adsorbed magnetic impurities in graphene:
  Symmetry and strain effects.
\newblock {\em Physical Review B}, 88(8):085405, August 2013.

\bibitem{sun_indirect_2013}
J.~H. Sun, F.~M. Hu, H.~K. Tang, W.~Guo, and H.~Q. Lin.
\newblock Indirect exchange of magnetic impurities in zigzag graphene ribbon.
\newblock {\em Journal of Applied Physics}, 113(17):17B515--17B515--3, March
  2013.

\bibitem{soriano_hydrogenated_2010}
D.~Soriano, F.~Mu\~{n}oz Rojas, J.~Fern\'{a}ndez-Rossier, and J.~J. Palacios.
\newblock Hydrogenated graphene nanoribbons for spintronics.
\newblock {\em Physical Review B}, 81(16):165409, April 2010.

\bibitem{bunder_ruderman-kittel-kasuya-yosida_2009}
J.~E. Bunder and Hsiu-Hau Lin.
\newblock Ruderman-kittel-kasuya-yosida interactions on a bipartite lattice.
\newblock {\em Physical Review B}, 80(15):153414, October 2009.

\bibitem{szalowski_indirect_2013}
Karol Sza\l{}owski.
\newblock Indirect {RKKY} interaction between localized magnetic moments in
  armchair graphene nanoribbons.
\newblock {\em Journal of Physics: Condensed Matter}, 25(16):166001, April
  2013.

\bibitem{acik_nature_2011}
Muge Acik and Yves~J. Chabal.
\newblock Nature of graphene edges: A review.
\newblock {\em Japanese Journal of Applied Physics}, 50(7):070101, 2011.

\bibitem{girit_graphene_2009}
\c{C}a\v{g}lar~\"{O} Girit, Jannik~C. Meyer, Rolf Erni, Marta~D. Rossell,
  C.~Kisielowski, Li~Yang, Cheol-Hwan Park, M.~F. Crommie, Marvin~L. Cohen,
  Steven~G. Louie, and A.~Zettl.
\newblock Graphene at the edge: Stability and dynamics.
\newblock {\em Science}, 323(5922):1705--1708, March 2009.
\newblock {PMID:} 19325110.

\bibitem{son_energy_2006}
Young-Woo Son, Marvin~L. Cohen, and Steven~G. Louie.
\newblock Energy gaps in graphene nanoribbons.
\newblock {\em Physical Review Letters}, 97(21):216803, November 2006.

\bibitem{han_energy_2007}
Melinda~Y. Han, Barbaros \"{O}zyilmaz, Yuanbo Zhang, and Philip Kim.
\newblock Energy band-gap engineering of graphene nanoribbons.
\newblock {\em Physical Review Letters}, 98(20):206805, May 2007.

\bibitem{power_model_2009}
S.~R. Power, V.~M. de~Menezes, S.~B. Fagan, and M.~S. Ferreira.
\newblock Model of impurity segregation in graphene nanoribbons.
\newblock {\em Physical Review B}, 80(23):235424, December 2009.

\bibitem{klinovaja_rkky_2013}
Jelena Klinovaja and Daniel Loss.
\newblock {RKKY} interaction in carbon nanotubes and graphene nanoribbons.
\newblock {\em Physical Review B}, 87(4):045422, January 2013.

\bibitem{fujita_peculiar_1996}
Mitsutaka Fujita, Katsunori Wakabayashi, Kyoko Nakada, and Koichi Kusakabe.
\newblock Peculiar localized state at zigzag graphite edge.
\newblock {\em Journal of the Physical Society of Japan}, 65(7):1920--1923,
  1996.

\bibitem{nakada_edge_1996}
Kyoko Nakada, Mitsutaka Fujita, Gene Dresselhaus, and Mildred~S. Dresselhaus.
\newblock Edge state in graphene ribbons: Nanometer size effect and edge shape
  dependence.
\newblock {\em Physical Review B}, 54(24):17954--17961, December 1996.

\bibitem{kunstmann_stability_2011}
Jens Kunstmann, Cem \"{O}zdo\c{g}an, Alexander Quandt, and Holger Fehske.
\newblock Stability of edge states and edge magnetism in graphene nanoribbons.
\newblock {\em Physical Review B}, 83(4):045414, January 2011.

\bibitem{son_half-metallic_2006}
Young-Woo Son, Marvin~L. Cohen, and Steven~G. Louie.
\newblock Half-metallic graphene nanoribbons.
\newblock {\em Nature}, 444(7117):347--349, November 2006.

\bibitem{yazyev_magnetic_2008}
Oleg~V. Yazyev and M.~I. Katsnelson.
\newblock Magnetic correlations at graphene edges: Basis for novel spintronics
  devices.
\newblock {\em Physical Review Letters}, 100(4):047209, January 2008.

\bibitem{wang_z-shaped_2007}
Z.~F. Wang, Q.~W. Shi, Qunxiang Li, Xiaoping Wang, J.~G. Hou, Huaixiu Zheng,
  Yao Yao, and Jie Chen.
\newblock Z-shaped graphene nanoribbon quantum dot device.
\newblock {\em Applied Physics Letters}, 91(5):053109--053109--3, July 2007.

\bibitem{zhang_experimentally_2012}
Xiaowei Zhang, Oleg~V. Yazyev, Juanjuan Feng, Liming Xie, Chenggang Tao,
  Yen-Chia Chen, Liying Jiao, Zahra Pedramrazi, Alex Zettl, Steven~G. Louie,
  Hongjie Dai, and Michael~F. Crommie.
\newblock Experimentally engineering the edge termination of graphene
  nanoribbons.
\newblock {\em {ACS} Nano}, November 2012.

\bibitem{wang_etching_2010}
Xinran Wang and Hongjie Dai.
\newblock Etching and narrowing of graphene from the edges.
\newblock {\em Nature Chemistry}, 2(8):661--665, 2010.

\bibitem{kosynkin_longitudinal_2009}
Dmitry~V. Kosynkin, Amanda~L. Higginbotham, Alexander Sinitskii, Jay~R. Lomeda,
  Ayrat Dimiev, B.~Katherine Price, and James~M. Tour.
\newblock Longitudinal unzipping of carbon nanotubes to form graphene
  nanoribbons.
\newblock {\em Nature}, 458(7240):872--876, April 2009.

\bibitem{lloyd_wave_1967}
P~Lloyd.
\newblock Wave propagation through an assembly of spheres: {II.} the density of
  single-particle eigenstates.
\newblock {\em Proceedings of the Physical Society}, 90(1):207--216, January
  1967.

\bibitem{ferreira_fundamental_1996}
M~S Ferreira, J~{d'Albuquerque}~e Castro, D~M Edwards, and J~Mathon.
\newblock Fundamental oscillation periods of the interlayer exchange coupling
  beyond the {RKKY} approximation.
\newblock {\em Journal of Physics: Condensed Matter}, 8(50):11259--11276,
  December 1996.

\bibitem{sancho_quick_1984}
M~P~Lopez Sancho, J~M~Lopez Sancho, and J~Rubio.
\newblock Quick iterative scheme for the calculation of transfer matrices:
  application to {Mo} (100).
\newblock {\em Journal of Physics F: Metal Physics}, 14(5):1205--1215, May
  1984.

\bibitem{kirwan_sudden_2008}
D.~F. Kirwan, C.~G. Rocha, A.~T. Costa, and M.~S. Ferreira.
\newblock Sudden decay of indirect exchange coupling between magnetic atoms on
  carbon nanotubes.
\newblock {\em Physical Review B}, 77(8):085432, February 2008.

\bibitem{costa_indirect_2005}
A.~T. Costa, D.~F. Kirwan, and M.~S. Ferreira.
\newblock Indirect exchange coupling between magnetic adatoms in carbon
  nanotubes.
\newblock {\em Physical Review B}, 72(8):085402, August 2005.

\bibitem{heinrich_dynamic_2003}
Bret Heinrich, Yaroslav Tserkovnyak, Georg Woltersdorf, Arne Brataas, Radovan
  Urban, and Gerrit E.~W. Bauer.
\newblock Dynamic exchange coupling in magnetic bilayers.
\newblock {\em Physical Review Letters}, 90(18):187601, May 2003.

\bibitem{guimaraes_spin_2011}
F.~S.~M. Guimarães, A.~T. Costa, R.~B. Muniz, and D.~L. Mills.
\newblock Spin currents in metallic nanostructures: Explicit calculations.
\newblock {\em Physical Review B}, 84(5):054403, August 2011.

\end{thebibliography}
\bibliographystyle{unsrt}

\end{document}